\documentclass[10pt, conference]{IEEEtran}
\IEEEoverridecommandlockouts
\usepackage{todonotes}
\usepackage{tikz}
\usepackage{hyperref}
\usetikzlibrary{quantikz2}
\usepackage{float}
\usepackage{url}
\usepackage[thinlines]{easytable}
\usepackage{comment}
\usepackage[n,
 advantage,
 operators,
 sets,
 adversary,
 landau,
 probability,
 notions,
 logic,
 ff,
 mm,
 primitives,
 events,
 complexity,
 oracles,
 asymptotics,
 keys] {cryptocode}
\usepackage{tikzit}

\tikzstyle{X Spider}=[fill={rgb,255: red,232; green,165; blue,165}, draw=black, shape=circle, tikzit fill={rgb,255: red,232; green,165; blue,165}, tikzit draw=black]
\tikzstyle{Z Spider}=[fill={rgb,255: red,216; green,248; blue,216}, draw=black, shape=circle, tikzit fill={rgb,255: red,216; green,248; blue,216}, tikzit draw=black]
\tikzstyle{Hadamard}=[fill=yellow, draw=black, shape=rectangle, tikzit draw=black, tikzit fill=yellow]

\usepackage{cite}
\usepackage{amsmath,amssymb,amsfonts}
\usepackage{algorithmic}
\usepackage{graphicx}
\usepackage{textcomp}
\usepackage{xcolor}
\usepackage{mathtools}

\begin{document}

\title{Optimizing State Preparation for Variational Quantum Regression on NISQ Hardware\\
\thanks{This work was funded by the Business Finland Quantum Computing
Campaign: projects FrameQ and EM4QS.}
}

\author{\IEEEauthorblockN{1\textsuperscript{st} Frans Perkkola}
\IEEEauthorblockA{\textit{Dept. of Computer Science} \\
\textit{University of Helsinki}\\
Helsinki, Finland \\
frans.perkkola@helsinki.fi}
\and
\IEEEauthorblockN{2\textsuperscript{nd} Ilmo Salmeperä}
\IEEEauthorblockA{\textit{Dept. of Computer Science} \\
\textit{University of Helsinki}\\
Helsinki, Finland \\
ilmo.salmenpera@helsinki.fi}
\and
\IEEEauthorblockN{3\textsuperscript{rd} Arianne Meijer-van de Griend}
\IEEEauthorblockA{\textit{Dept. of Computer Science} \\
\textit{University of Helsinki}\\
Helsinki, Finland \\
arianne.vandegriend@helsinki.fi}
\and
\IEEEauthorblockN{4\textsuperscript{th} C.-C. Joseph Wang}
\IEEEauthorblockA{\textit{Quantum Computational Science Group} \\
\textit{Oak Ridge National Laboratory}\\
Oak Ridge, USA \\
wangccj@ornl.gov}
\and
\IEEEauthorblockN{5\textsuperscript{th} Ryan S. Bennink}
\IEEEauthorblockA{\textit{Quantum Computational Science Group} \\
\textit{Oak Ridge National Laboratory}\\
Oak Ridge, USA \\
benninkrs@ornl.gov}
\and
\IEEEauthorblockN{6\textsuperscript{th} Jukka K. Nurminen}
\IEEEauthorblockA{\textit{Dept. of Computer Science} \\
\textit{University of Helsinki}\\
Helsinki, Finland \\
jukka.k.nurminen@helsinki.fi}
}

\maketitle

\begin{abstract}
The execution of quantum algorithms on modern hardware is often constrained by noise and qubit decoherence, limiting the circuit depth and the number of gates that can be executed. Circuit optimization techniques help mitigate these limitations, enhancing algorithm feasibility. In this work, we implement, optimize, and execute a variational quantum regression algorithm using a novel state preparation method. By leveraging ZX-calculus-based optimization techniques, such as Pauli pushing, phase folding, and Hadamard pushing, we achieve a more efficient circuit design. Our results demonstrate that these optimizations enable the successful execution of the quantum regression algorithm on current hardware. Furthermore, the techniques presented are broadly applicable to other quantum circuits requiring arbitrary real-valued state preparation, advancing the practical implementation of quantum algorithms.
\end{abstract}


\section{Introduction}
Executing quantum algorithms on modern quantum hardware is difficult because of the errors introduced by imperfect control over the qubits. These errors restrict the circuit depth, often limiting us to fewer gates than a straightforward implementation of an algorithm requires. Therefore, circuit optimization is essential to ensure we get reasonable results from a quantum computer instead of random noise. Modern quantum compilers provide some standard circuit optimization techniques, but these techniques can be insufficient. Therefore, manual optimization is needed to run the experiments.

In this article, we implement, optimize and execute the variational quantum regression algorithm by Wang and Bennink \cite{wang2024variationalquantumregressionalgorithm} on real quantum hardware. Their approach uses the amplitude encoding algorithm as a feature map to map real-valued classical data points onto the probability amplitudes of the quantum superposition states. The ansatz of the algorithm then applies the regression coefficients directly to the corresponding features as variational parameters. Therefore, the output of the quantum circuit after the measurements is the mean squared error of the weighted data points. The key difference between this algorithm and many other quantum machine learning algorithms is that this algorithm processes multiple data points simultaneously instead of one at a time.

The main bottleneck of the variational quantum regression algorithm is the quantum state preparation. It is well known that arbitrary quantum state preparation without using multiple ancilla qubits is an expensive subroutine, as the number of elementary gates needed for it grows exponentially with respect to the number of qubits involved. Therefore, the subroutine must be optimized as much as possible. Current state-of-the-art state preparation algorithms are at best 2-approximations of the optimized state preparation algorithm introduced in this article when \textit{all} elementary operations are considered \cite{PhysRevA.93.032318, mottonen2004transformationquantumstatesusing}. 

The main circuit optimization techniques used in this article are well-known methods from ZX-calculus: Pauli pushing, phase folding, and Hadamard pushing \cite{vandewetering2020zxcalculusworkingquantumcomputer, amy2024linearnonlinearrelationalanalyses}. The gate-pushing techniques involve pushing gates through the quantum circuit following Clifford commutation rules. The phase folding technique allows us to sum up the components of the decomposed multi-controlled $R_Z$ gates. The key contribution of this article is to demonstrate how these optimization techniques enable us to transform an algorithm that, in its naive form, exceeds the capabilities of current quantum computers. By applying circuit optimization techniques effectively, we can significantly shorten the algorithm, making it feasible to execute.

In addition to circuit optimization, it is practical to use classical post-processing techniques to enhance the performance of the quantum algorithm. In this work, we use and analyze an error mitigation technique \cite{PRXQuantum.2.040326} to mitigate measurement errors and a classical shadow technique \cite{Huang_2020}, which enables us to estimate the expected value of a quantum circuit more accurately.

The circuit optimization techniques in this article applied to the state preparation method introduced in \cite{wang2024variationalquantumregressionalgorithm} yields a new, more optimized method for preparing arbitrary quantum states. Furthermore, using this state preparation method as a subroutine in the quantum regression algorithm also proposed in \cite{wang2024variationalquantumregressionalgorithm} produces a quantum regression algorithm usable on current quantum hardware. Even though this article focuses on the regression algorithm presented in \cite{wang2024variationalquantumregressionalgorithm}, these optimization techniques could be applied to any quantum circuit with a similar structure, such as Variational Quantum Algorithms. Furthermore, the new state preparation method can be used as a subroutine for other quantum algorithms that require arbitrary state preparation for real-valued amplitudes.

The key contributions of this article are:
\begin{itemize}
    \item Demonstrating the successful execution of a quantum regression algorithm on real quantum hardware.
    \item Presenting circuit optimization techniques usable also for other quantum algorithms
    \item Introducing an optimized state preparation method for real-valued amplitudes
    \item Demonstrating the effects of classical post-processing techniques for the model accuracy
\end{itemize}

\section{Background}
This section briefly describes the quantum regression algorithm proposed in \cite{wang2024variationalquantumregressionalgorithm} which we practically demonstrate on the quantum computer. 
The quantum regression algorithm was proposed using two distinct encoding methods: the compact binary encoding and one-hot encoding.
This article focuses solely on the compact binary encoding method, because the one-hot encoding method exceeds the qubit and quantum gate limitations of modern quantum hardware and is thus not feasible for real-world use cases at the time of writing.
The compact binary method involves three steps: state preparation, a quantum regression map, and a measurement of an observable. In this section, we will briefly explain each step.

\subsection{State preparation}
The state preparation in the compact binary method begins by preparing a uniform superposition of $K=L(M+1)$ states and an ancilla qubit in the $|+\rangle$ state:
\begin{equation}
    |\Psi \rangle = |+\rangle \otimes (\frac{1}{\sqrt{K}} \sum_k |k\rangle),
\end{equation}
where $L$ denotes the number of rows, and $M+1$ is the number of columns in the data table.

After that, we use multi-controlled $R_Z$ gates together with X gates to impart a desired phase $x_k$ on each state:
\begin{equation}
    U_D^k = \exp(-ix_kZ_A\otimes|k\rangle\langle k|).
\end{equation}

Applying the unitary $U_D^k$ for all $k$ on the uniform state $|\Psi\rangle$ yields the following state:
\begin{equation}
    \label{prod}
    \prod_k U_D^k |\Psi\rangle = \frac{1}{\sqrt{K}} \sum_k \frac{e^{-ix_k} |0\rangle+e^{ix_k} |1\rangle}{\sqrt{2}} \otimes | k \rangle.
\end{equation}

Finally, the encoded data state is realized by projecting the ancilla qubit onto the $|-\rangle$ state:

\begin{equation}
    \langle - | \prod_k U_D^k |\Psi\rangle \propto \sum_k \sin x_k |k\rangle \approx \sum_k x_k |k\rangle = |\psi_D\rangle,
\end{equation}
since $\sin x_k \approx x_k$ for a normalized data table i.e. small values of $x_k$.

\subsection{The quantum regression map}
To impart the regression coefficients into the encoded data state $|\psi_D\rangle$, we will use a gadget similar to $U_D^k$:

\begin{equation}
    U_C^m = \exp(i\phi_mZ_A\otimes \boldsymbol{1}\otimes |m\rangle\langle m |).
\end{equation}

The main difference is that where $U_D^k$ selects a specific data element $k = (l, m)$, $U_C^m$ selects only the column $m$, and performs identically on each row $(l)$ of the encoded data table.

Again, we start by preparing an ancilla qubit in the $|+\rangle$ state:
\begin{equation}
    |\Psi_C\rangle = |+\rangle \otimes |\psi_D\rangle.
\end{equation}

After that, we impart the regression coefficients for each $m$ using the $U_C^m$ gadget:

\begin{equation}
    \prod_m U_C^m |\Psi_C\rangle = \sum_{l,m}\frac{e^{i\phi_m}|0\rangle + e^{-i\phi_m} |1\rangle}{\sqrt{2}} \otimes x_{lm} |lm\rangle.
\end{equation}

Finally, projecting the ancilla qubit onto the $|+\rangle$ state yields the final quantum state:

\begin{equation}
    \langle + |\prod_m U_C^m |\Psi_C\rangle = \sum_{l,m} x_{lm} \cos \phi_m |lm\rangle = |\Psi_0\rangle
\end{equation}

\subsection{Measurement}
The prepared quantum state $|\Psi_0\rangle$ represents a data table with regression weights $\cos \phi_m$ for each column $m$. Define the measurement operator as

\begin{equation}
    \hat{M} = I^{\otimes N_L}\otimes(I+X)^{\otimes N_M},
\end{equation}
where $N_L$ is defined as $\ceil{\log L}$ and $N_M$ as $\ceil{\log(M+1)}$.

Using this, we can calculate the expected value of $\hat{M}$ with respect to the quantum state $|\Psi_0\rangle$:

\begin{equation}
\label{expected}
    \langle \hat{M} \rangle = (\cos\phi_0)^2\sum_l(y_l-\sum_{m=1}^Mx_{lm}W_m)^2,
\end{equation}
where we identify $W_m=-\cos \phi_m /\cos \phi_0$ as the regression coefficient for feature $m$ with $M$ features, and the response variable $y_l$ is the $x_{l0}$ component \cite{wang2024variationalquantumregressionalgorithm}. The expected value in (\ref{expected}) can thus be directly used as a loss function in the variational quantum algorithm.

\section{Constructing the unitaries}
The naive implementation of the $U_D$ and $U_C$ unitaries requires $K+M+1$ multi-controlled $R_Z$ gates together with $\approx 2K\log K + 2(M+1) \log (M+1)$ X gates and $2\log (K+2)$ Hadamard gates. The X gates are used to realize the $|k\rangle \langle k|$ and $|m\rangle \langle m |$ in the definitions of $U_D^k$ and $U_C^m$, respectively. An example of this can be found in Fig. \ref{fig:optimized}a.

The multi-controlled $R_Z$ gates are not native to any quantum hardware, so we must decompose them into elementary gates. However, this approach leads to an inefficient method for realizing the said unitaries as decomposing multi-controlled gates leads to an exponential amount of elementary gates with respect to the number of control qubits \cite{Vartiainen_2004}. In this article, we use the methods of Möttönen, Vartiainen, Bergholm, and Salomaa \cite{mottonen2004transformationquantumstatesusing} to decompose the multi-controlled unitaries.

\begin{figure*}[t]
    \centering
    \[a)
    \begin{quantikz}[row sep={0.6cm,between origins}, column sep = 0.1cm]
        & \ctrl{3} & \\
        & \ctrl{2} & \\
        & \ctrl{1} & \\
        & \gate{R_Z(\theta)} &
    \end{quantikz} =  \begin{quantikz}[row sep={0.6cm,between origins}, column sep = 0.1cm]
        & &\ctrl{3} & & & & \ctrl{3} & & & & \ctrl{3} & & & & \ctrl{3} & & &\\
        & & & & \ctrl{2} & & & & & & & & \ctrl{2} & & & & &\\
        & & & & & & & & \ctrl{1} & & & & & & & & \ctrl{1} &\\
        & \gate{R_Z(\frac{\theta}{8})} & \targ{} & \gate{R_Z(\frac{-\theta}{8})} & \targ{} & \gate{R_Z(\frac{\theta}{8})} & \targ{} & \gate{R_Z(\frac{-\theta}{8})} & \targ{} & \gate{R_Z(\frac{\theta}{8})} & \targ{} & \gate{R_Z(\frac{-\theta}{8})} & \targ{} & \gate{R_Z(\frac{\theta}{8})} & \targ{} & \gate{R_Z(\frac{-\theta}{8})} & \targ{} &
    \end{quantikz}
    \]\[b)
     \begin{quantikz}[row sep={0.6cm,between origins}, column sep = 0.1cm]
        & & \ctrl{2} & & & & \ctrl{2} & & & & \ctrl{2} & & & & \ctrl{2} & & &\\
        & & & & \ctrl{1} & & & & \ctrl{1} \slice{} & & & & \ctrl{1} & & & & \ctrl{1} &\\
        & \gate{R_Z(\frac{-\theta_i}{4})}  & \targ{} & \gate{R_Z(\frac{-\theta_i}{4})} & \targ{} & \gate{R_Z(\frac{\theta_i}{4})} & \targ{} & \gate{R_Z(\frac{\theta_i}{4})} & \targ{} & \gate{R_Z(\frac{\theta_j}{4})} & \targ{} & \gate{R_Z(\frac{-\theta_j}{4})} & \targ{} & \gate{R_Z(\frac{\theta_j}{4})} & \targ{} & \gate{R_Z(\frac{-\theta_j}{4})} & \targ{} &
    \end{quantikz}\] 
    =\begin{quantikz}[row sep={0.6cm,between origins}, column sep = 0.1cm]
        & & \ctrl{2} & & & & \ctrl{2} & & &  & & & & & & \\
        & & & & \ctrl{1} & & & & \ctrl{1} & & \\
        & \gate{R_Z(\frac{\theta_{j}-\theta_{i}}{4})} &\targ{} & \gate{R_Z(-\frac{(\theta_{i}+\theta_{j})}{4})} & \targ{} & \gate{R_Z(\frac{\theta_{i}+\theta_{j}}{4})}& \targ{} & \gate{R_Z(\frac{(\theta_{i}-\theta_{j})}{4})}& \targ{} &
    \end{quantikz}
    \caption{Decomposing and optimizing multi-controlled $R_Z$ gates. a) Decomposition of a 3-controlled $R_Z$ gate. b) A simple example of phase folding.}
    \label{fig:3-control}
\end{figure*}
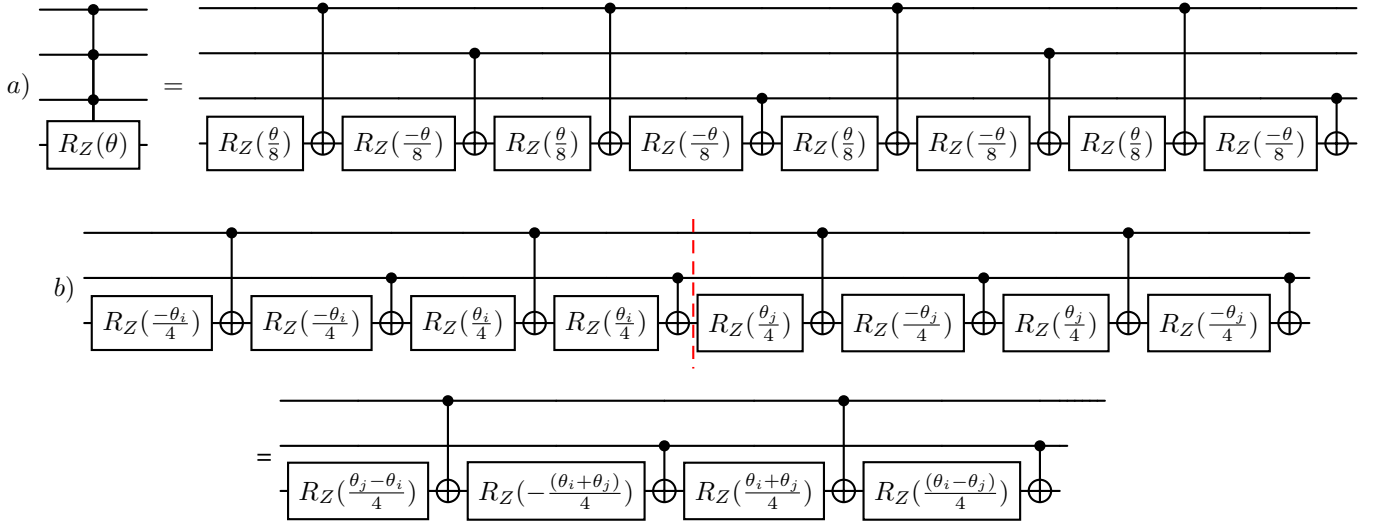
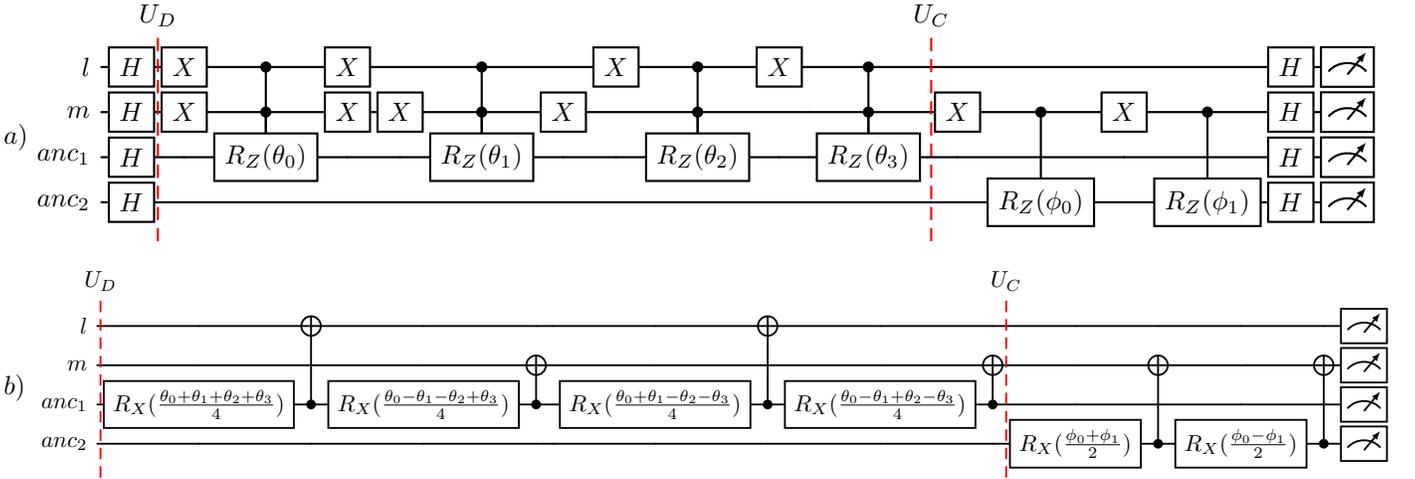
\begin{figure*}[t]
    \centering
    \[a) \begin{quantikz}[row sep={0.6cm,between origins}, column sep = 0.1cm]
       \lstick{{$l$}} & \gate{H}\slice{{$U_D$}} &  \gate{X}  & \ctrl{2} & \gate{X} &   & \ctrl{2} &  & \gate{X} & \ctrl{2} & \gate{X} &  & \ctrl{2} & \slice{{$U_C$}} &  & & & & \gate{H} & \meter{}\\
       \lstick{{$m$}} & \gate{H} & \gate{X} & \ctrl{1} & \gate{X} &\gate{X} & \ctrl{1} & \gate{X} & & \ctrl{1} &  & & \ctrl{1} &  & \gate{X} & \ctrl{2} & \gate{X} & \ctrl{2} & \gate{H} & \meter{}\\
        \lstick{{$anc_1$}}& \gate{H} & & \gate{R_Z(\theta_0)} & && \gate{R_Z(\theta_1)} & && \gate{R_Z(\theta_2)} & && \gate{R_Z(\theta_3)} & & & & & & \gate{H} & \meter{}\\
       \lstick{{$anc_2$}} & \gate{H} & & & & & & & & & & & & & & \gate{R_Z(\phi_0)} & & \gate{R_Z(\phi_1)} & \gate{H} & \meter{}
    \end{quantikz}
    \]
    \[b)
    \scalebox{0.87}{    \begin{quantikz}[row sep={0.6cm,between origins}, column sep = 0.1cm]
        \lstick{{$l$}}\slice{{$U_D$}} & & \targ{}& & & & & \targ{} & &\slice{{$U_C$}} & & & & & \meter{}\\
        \lstick{{$m$}} & & & & \targ{} & & & & & \targ{} & & \targ{} & & \targ{} & \meter{}\\
        \lstick{{$anc_1$}} & \gate{R_X(\frac{\theta_0+\theta_1+\theta_2+\theta_3}{4})} & \ctrl{-2} & \gate{R_X(\frac{\theta_0-\theta_1-\theta_2+\theta_3}{4})} & \ctrl{-1} & & \gate{R_X(\frac{\theta_0+\theta_1-\theta_2-\theta_3}{4})} & \ctrl{-2} & \gate{R_X(\frac{\theta_0-\theta_1+\theta_2-\theta_3}{4})} & \ctrl{-1} & & & & &  \meter{}\\
        \lstick{{$anc_2$}} & & & & & & & & & & \gate{R_X(\frac{\phi_0 + \phi_1}{2})} & \ctrl{-2} & \gate{R_X(\frac{\phi_0 - \phi_1}{2})} & \ctrl{-2} & \meter{}
    \end{quantikz}}
    \]
    \caption{End-to-end example(s) for the optimization process for the quantum regression algorithm with the compact binary encoding. a) An initial, non-optimized circuit example with $K=4$ and $M=1$. Each of the multi-controlled $R_Z$ gates present decompose to an exponential amount of CNOT gates with respect to the amount of control qubits, similarly to the example in Fig. \ref{fig:3-control}a. b) An optimized version of the example circuit above. In this circuit, all-to-all qubit connectivity is assumed.}
    \label{fig:optimized}
\end{figure*}

As presented in \cite{mottonen2004transformationquantumstatesusing}, a $n$-controlled $R_Z$ gate is decomposed into $2^n$ CNOT gates and $2^n$ $R_Z$ gates. This results in an exponential complexity of $O(2^n)$ for a single n-controlled $R_Z$ gate. An example of this decomposition for a 3-controlled $R_Z$ gate can be seen in Fig. \ref{fig:3-control}a.

In the data preparation unitary $U_D$ we have $\log K$ control qubits and a total of $K$ data points to encode. Similarly, for the regression weight unitary $U_C$ we have $\log (M+1)$ control qubits and a total of $M+1$ weights to encode. Therefore, a naive approach for implementing $U_D$ together with $U_C$ gives us a total gate count of

\begin{equation}
\label{eq1}
\begin{split}
&2(\underbrace{K^2}_{U_D} + \underbrace{(M+1)^2}_{U_C}+\underbrace{K\log K + (M+1) \log (M+1)}_{\text{X gates}}\\
&+\underbrace{\log (K+2)}_{\text{H gates}})  
\end{split}
\end{equation}

 for the whole circuit when combining all CNOT, $R_Z$, X, and H gates, which is intractable for existing quantum hardware. Thus, we need to optimize this circuit. An example of the naive implementation can be seen in Fig. \ref{fig:optimized}a.

\section{Optimizing \texorpdfstring{$U_D$}{U_D} and \texorpdfstring{$U_C$}{U_C}}
\subsection{Pauli pushing}
We will start the optimization process by decomposing all of the multi-controlled $R_Z$ gates in the circuit. When they are in their decomposed form, we can optimize the circuit by removing the X gates present in the circuit using Pauli pushing. To do this, we can "push" the Pauli X gate through the circuit using Clifford commutation rules. Similarly, pushing the X gate through the circuit flips the sign of all the $R_Z$ gates it passes. Since X gates in $U_D$ and $U_C$ come in pairs, they cancel with each other, reducing the total gate count by $2K\log K+2(M+1) \log (M+1)$. Appendix \ref{ref:app_a} provides a step-by-step example.

\subsection{Phase folding}
After pushing the X gates, we optimize $U_D$ and $U_C$ using phase folding \cite{amy2024linearnonlinearrelationalanalyses}, a technique that essentially merges the decomposition of the multi-controlled $R_Z$ gates. 
As a result, we reduce the number of gates from $(K+M+1)G$ to only $2G$ where $G$ is the number of gates in a decomposed multi-controlled $R_Z$ gate. We can see a simple example of phase folding in Fig. \ref{fig:3-control}b, and Fig. \ref{fig:optimized}b shows its usage on an example circuit for the algorithm. Therefore, this technique optimizes the terms $2K^2$ and $2(M+1)^2$ in the total gate count to $2K$ and $2(M+1)$, respectively. Specifics on phase folding can be found in Appendix \ref{ref:app_b}.

\subsection{Hadamard pushing}
The last utilized optimization technique is called Hadamard pushing. The circuit has Hadamard gates at the beginning and end on all $\log (K + 2)$ qubits as shown in figure \ref{fig:optimized}a. Similarly to the Pauli X gates, Hadamard gates can be pushed to the end of the circuit where they cancel each other, turning $R_Z$ gates into $R_X$ gates and flipping the direction of CNOTs. This optimization method reduces the total gate count by $2\log (K+2)$. Specifics on Hadamard pushing can also be found in appendix \ref{ref:app_a}.

Summarizing, we started with a total gate count of $2(K^2 + (M+1)^2+K\log K + (M+1) \log (M+1)+\log (K+2))$ for the naive implementation, and we managed to optimize it to $2(K+M+1)$. Since almost always $K > M$, we get a final gate complexity of $O(K)$ after optimization. An example circuit for this algorithm and its optimized version can be found in Fig. \ref{fig:optimized}.

\section{Classical post-processing techniques }
One major problem with the hardware realization of the regression algorithm is that the projective measurement $\langle - |$  in the state preparation procedure succeeds with very low probability, leaving us with very few successful measurements to estimate the expected value with. To better estimate the expected value of the circuit, we used a classical shadow technique, which gives an approximate classical description
of the quantum state using very few measurements of the state \cite{Huang_2020}. Using the classical shadow technique involves a full measurement in the X-basis, which requires us to insert Hadamard gates at the end of the circuit as in Fig. \ref{fig:optimized}a.

Another problem with the hardware realization of the algorithm is measurement errors, which means that the measurement outcome is flipped with some probability due to hardware noise. To mitigate these errors we used the Mthree Python library \cite{PRXQuantum.2.040326}. The Mthree library first finds the physical qubits to which the virtual qubits are mapped. After that, it obtains their measurement noise profile. Finally, it applies the correction on the counts achieved from the circuit run. This mitigates the errors and allows for a slightly more accurate approximation of the expected value. 

\section{Experimental setup}
As the experimental dataset, we used Mohan S Acharya's Graduate Admission dataset \cite{aineisto}. The dataset consists of seven independent variables considered important in the application process of Indian students to Master's programs. The regression goal is to find the probability of student admission. The dataset contains 400 data points, of which 256 rows are selected as training data and the remaining 144 rows as test data. We used batching with a batch size of eight. The dataset was chosen as it is of suitable size and reasonable difficulty.

For the optimization of the regression weight parameters, we used the gradient descent method with the ADAM optimizer \cite{kingma2017adammethodstochasticoptimization}. For the learning rate, we used a value of 0.01. The ADAM optimizer was implemented by us in Python. For the gradient evaluation, we used the parameter-shift method. Therefore, one gradient evaluation required 16 circuit runs, that is, 2 circuit runs for each $M+1$ parameters. A total of eight qubits were used: two ancilla qubits and six data qubits.

This algorithm was executed on the IQM Garnet (Crystal 20) quantum machine \cite{garnet}. The IQM Garnet is a 20-qubit quantum processing unit based on superconducting transmon qubits. The qubits are arranged in a square lattice and connected by tunable couplers. At the time of execution, the machine had a 99.89\% one-qubit gate fidelity and 99.23\% two-qubit gate fidelity \cite{garnet}.

At the time of writing, the IQM Garnet does not allow mid-circuit measurements, so we had to use two ancilla qubits to perform the projective measurements on: one for $U_D$ and one for $U_C$. After the measurements, we used post-selection to capture the correctly projected states. For routing the logical qubits to physical qubits, we used Qiskit's built-in method, which uses the SABRE algorithm to route the qubits \cite{li2019tacklingqubitmappingproblem}. This approach proved to be sufficiently good, resulting in only a few dozen additional two- and one-qubit gates.

We trained the dataset on three different models: a classical regression model, a classically simulated model of the quantum algorithm without noise, and the hardware implementation model. The classical regression model has a classical mean squared error as the classically optimized loss function. The classically simulated model was simulated in the Qiskit environment using Qiskit's Aer simulator. The hardware model is the same as the noiseless model but runs on the IQM Garnet quantum device.

\section{Results}
\subsection{Effects of circuit optimization}
\begin{figure}[b]
    \centering
    \includegraphics[width=1\linewidth]{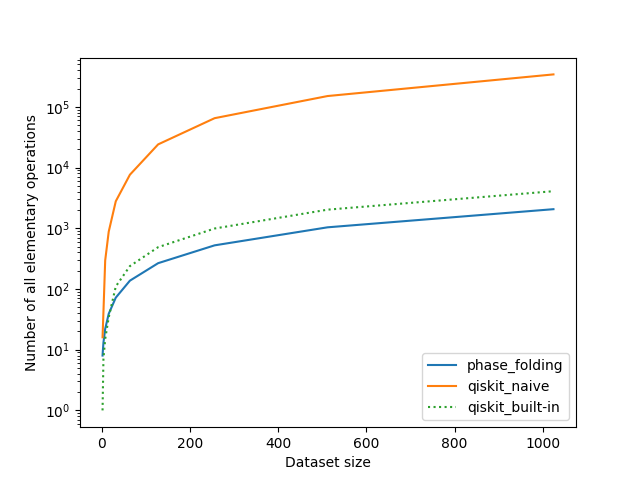}
    \caption{Scaling of all elementary gates in a naive implementation, optimized version, and Qiskit built-in state preparation.}
    \label{fig:benchmark}
\end{figure}

In Fig. \ref{fig:benchmark} we present the scaling of all elementary gates in the naive implementation of the state preparation part of this algorithm compared to the optimized version and Qiskit's built-in state preparation method. We notice that the naive implementation grows polynomially and the optimized version linearly with respect to the data table size $K$, as expected from our gate count analysis. The Qiskit's built-in state preparation method also grows linearly, but approximately twice as fast as our optimized method. The CNOT gate count is the same for the optimized and the Qiskit state preparation method. The plot was generated by preparing a state with random normalized, real-valued datasets of different sizes.

\begin{figure}[!t]
\raggedright
$a)$
 \includegraphics[width=1\linewidth]{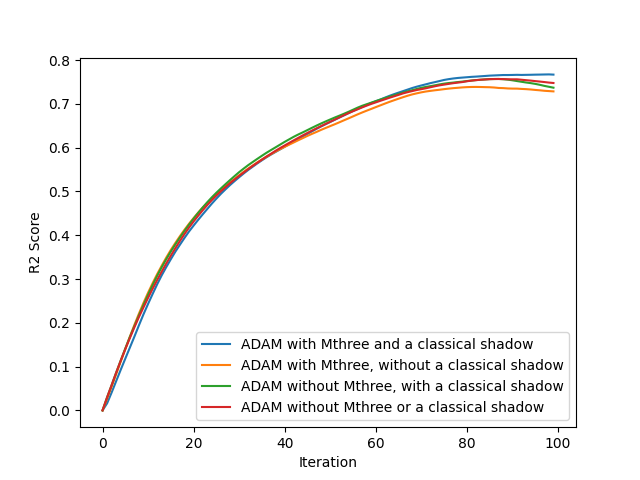}
$b)$
\includegraphics[width=1\linewidth]{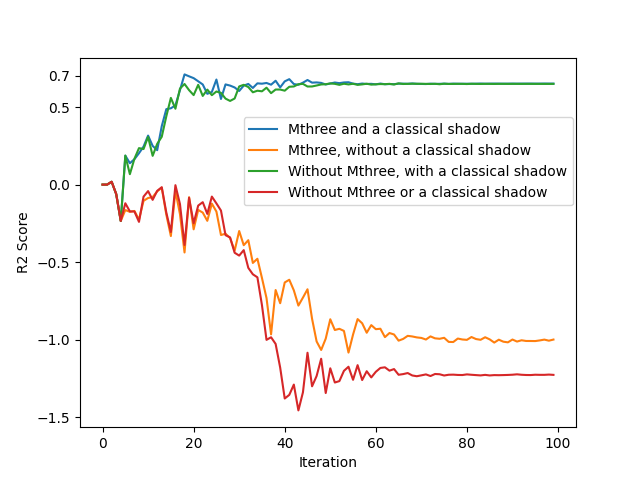}
    \caption{The effects of the classical post-processing techniques on the R2 score. a) ADAM with and without MThree and the classical shadow. b) Nelder-Mead with and without MThree and the classical shadow.}
    \label{fig:cl-post}
\end{figure}

\subsection{Effects of classical post-processing}
Fig. \ref{fig:cl-post} shows the effects of the classical post-processing techniques on the R2 score of the model. The R2 score, also known as the coefficient of determination, describes how well the predictions of a regression model match the actual data. It ranges from 1 (perfect fit) to 0 (no explanatory power) and can be negative if the model is worse than a simple mean prediction. 

We ran simulations of the quantum algorithm using the error model of the IQM Garnet (Crystal 20) quantum machine. The simulations were run with and without the MThree error mitigation and the classical shadow technique, taking the average over five runs for each simulation. A generated dataset of 64 data points was used, and the optimization was done with the Nelder-Mead (Fig. \ref{fig:cl-post}b) and ADAM (Fig. \ref{fig:cl-post}a) optimizers.

It is clear from Fig. \ref{fig:cl-post} that the classical shadow technique is essential for achieving any reasonable results when using the Nelder-Mead optimizer. The effects of the error mitigation are not as drastic, but become more relevant when running the algorithm on real quantum hardware. Similarly, we plotted the results with the ADAM optimizer, with and without error mitigation and the classical shadow. The figure shows, that the classical post-processing techniques are not as impactful as with the Nelder-Mead optimizer. However, it is reasonable to assume that these methods become more relevant when running the algorithm on real hardware. In this example, the ADAM optimizer achieved an R2 score of 0.767, while the Nelder-Mead reached only 0.710 at best. Both optimizers were run for 100 iterations and both optimizers ran the same amount of circuits per iteration.

\subsection{Transpilation and routing}
\label{routing}
In our hardware experiments, 72 RX gates and 72 CNOT gates were used for each quantum circuit. When transpiled and routed to the IQM Garnet hardware using the SABRE algorithm, these gates were mapped to approximately 130 R gates and 90 CZ gates, which are native to the hardware. Of these, approximately half of the R gates were mapped to the same qubit and the rest were distributed evenly for the rest of the qubits. The CZ gates were distributed evenly between the ancilla qubits and other non-ancilla qubits.

\subsection{Hardware results}
\begin{figure*}[t]
    \centering
    \includegraphics[width=0.8\linewidth]{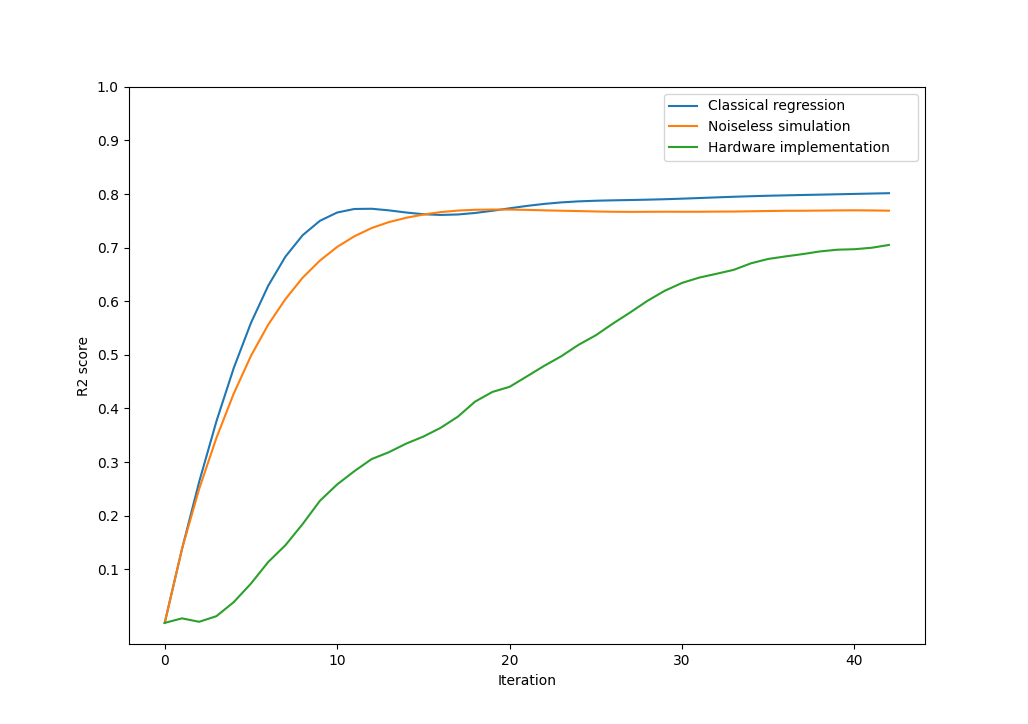}
    \caption{R2 score results for the Admission Predict dataset with the IQM Garnet quantum hardware, a noiseless simulator, and a purely classical regressor.}
    \label{fig:hardware-results}
\end{figure*}

Fig. \ref{fig:hardware-results} shows the R2 scores for a purely classical model of the algorithm compared to a noiseless quantum simulation together with the hardware implementation plotted for each iteration of the optimization. The purely classical model outperformed the simulated and the hardware models with the highest R2 score of 0.802. The simulated quantum algorithm achieved an R2 score of 0.769, while the hardware implementation reached 0.705. All models used ADAM as the optimizer.

Table \ref{tab:weights} shows the feature weights obtained from each of the models for a standardized data table. The weights from the classical regression model can be thought of as the baseline for the other models. Table \ref{tab:distances} shows the distances of the hardware implementation and the noiseless simulation feature weights from the baseline. The classical regression, the hardware implementation, and the noiseless simulation weights are denoted as $\widetilde{W}_{\text{cl-reg}}$, $\widetilde{W}_{\text{hardware}}$, and $\widetilde{W}_{\text{noiseless-sim}}$, respectively.

In the case of the quantum models, the circuits were sampled 20000 times. This took, on average, eight seconds of wall clock time on the IQM Garnet quantum device, not including the time spent on sending data back and forth with the IQM cloud. The total runtime of the hardware implementation was approximately three hours. Most of the time was spent evaluating the gradients, since one gradient evaluation required 16 circuit runs. The initialization experiments in MThree also took a significant amount of time.

\begin{table*}[t]
    \centering
    \caption{Feature weights obtained by the different models on the Admission Predict dataset.}
    \begin{center}
        
    \begin{tabular}{|c|c|c|c|c|c|c|c|}
    \hline
     & \small$\widetilde{W_1}$ & \small$\widetilde{W_2}$ & \small$\widetilde{W_3}$ & \small$\widetilde{W_4}$ & \small$\widetilde{W_5}$ & \small$\widetilde{W_6}$ & \small$\widetilde{W_7}$ \\ 
     \hline
    Classical regression & 0.05448049 & 0.34598482 & 0.11552738 & 0.02393412 & 0.07771921 & 0.1704666 & 0.23534216 \\
    \hline
    Hardware implementation & 0.1386957  & 0.19968415 & 0.06301308 & 0.05891065 & 0.06662064 & 0.1224363 &
 0.0598611 \\
    \hline
    Noiseless simulation & 0.16960455 & 0.20276255 & 0.14055369 & 0.16670063 & 0.1455859 &  0.12953202 &
 0.22334538 \\
    \hline
    \end{tabular}
    \label{tab:weights}
    \end{center}
\end{table*}

\begin{table*}[t]
    \centering
    \caption{Deviation of the feature weights of the hardware implementation and the noiseless classical simulation with respect to the classical regression as a distance.}
    \begin{center}
    \begin{tabular}{|c|c|c|c|c|c|c|c|c|}
    \hline
     & \small$\widetilde{W_1}$ & \small$\widetilde{W_2}$ & \small$\widetilde{W_3}$ & \small$\widetilde{W_4}$ & \small$\widetilde{W_5}$ & \small$\widetilde{W_6}$ & \small$\widetilde{W_7}$ \\ 
     \hline
     $|\widetilde{W}_{\text{cl-reg}}-\widetilde{W}_{\text{hardware}}|$ & 0.08421521 & 0.14630067 & 0.0525143 &  0.03497653 & 0.01109857 & 0.0480303 &
 0.17548106 \\
     \hline
      $|\widetilde{W}_{\text{cl-reg}}-\widetilde{W}_{\text{noiseless-sim}}|$ & 0.11512406 & 0.14322227 & 0.02502631 & 0.14276651 & 0.06786668 & 0.04093459 &
 0.01199678 \\
      \hline
    \end{tabular}
    \label{tab:distances}
    \end{center}
\end{table*}

\section{Discussion}
Out of the three models presented in Fig. \ref{fig:hardware-results}, the purely classical model is clearly the best. However, even in the presence of noise, the hardware model was able to learn and achieve results comparable to the classical model. Although the runtime on the hardware was long, the quantum algorithm could theoretically outperform classical linear regression, with a widely accepted time complexity of $O(KM)$ for the regression map due to slightly better time complexity scaling of $O(K)$ of the quantum algorithm. Additionally, there is some classical overhead included in the runtime that could be optimized, for example, the data normalization and the quantum circuit preparation. Furthermore, the classical communication between the devices may be a bottleneck. Therefore, we should not draw conclusions solely based on the wall clock runtime of the quantum algorithm.

The regression weights obtained by the hardware model do not deviate much from those of the classical regression baseline, as can be seen in tables \ref{tab:weights} and \ref{tab:distances}. Similarly, the weights obtained by the noiseless simulation appear to be close to the baseline as well. Between the hardware model and the noiseless simulation, some weights are closer to the baseline than others. This result can be explained by realizing that the solutions to the regression problem obtained by loss function optimization techniques are not always unique.

One major drawback of this algorithm is the fact that the projective measurement to the state $|-\rangle$ succeeds with very low probability. This happens because, in the state preparation part, we assume $\sum_k \sin^2(x_k)\approx1$. However, the probability to measure a 1 on the ancilla qubit is given by $\sum_k \sin^2(x_k) / K\approx1/K$. In other words, the probability of the projective measurement succeeding is inversely proportional to the data table size $K$. In our hardware experiments, this probability was $1/64$. Fixing this issue would unlock the full potential of this algorithm, and possibly present a new state-of-the-art state preparation method.

In the hardware experiment with the real-world dataset, we chose to use the ADAM optimizer over the Nelder-Mead optimizer used in the original article \cite{wang2024variationalquantumregressionalgorithm}, since the ADAM optimizer performed better when simulating the algorithm with an error model based on the targeted hardware device as can be seen in Fig. \ref{fig:cl-post}. Furthermore, when using the Nelder-Mead optimizer with the Graduate Admission dataset, the model was unable to learn anything. In contrast, the ADAM optimizer had a more stable performance, giving less variance in accuracy between different runs with a decent R2 score while simulating the algorithm with the target hardware noise model.

The gradient-based ADAM optimizer appeared to perform better than the Nelder-Mead optimizer in the absence of correctly projected states. This is likely because Nelder-Mead relies only on the loss function values, whereas ADAM moves in a direction defined by the gradient. Therefore, a significant amount of noise and a poor estimate of the expected value (loss value) can cause the Nelder-Mead to move in a random direction. In the case of ADAM, even with noise and few samples to calculate the expected value with, the gradient will likely be in the generally correct direction since the gradient is calculated for each parameter separately. 

Even with the help of the classical shadow, we were not able to estimate the expected value even to an accuracy of one decimal place consistently when using the Nelder-Mead optimizer on the Graduate Admission dataset, which is why we decided to use the ADAM optimizer. Fixing the issue with the projective measurement could also make the Nelder-Mead a viable method.

Using the decomposition in Fig. \ref{fig:3-control} for the multi-controlled gates is efficient in terms of CNOT count but it has one drawback, namely, the fact that most of the single-qubit gates are applied on the same qubit. As mentioned in section \ref{routing}, approximately half of the single qubit gates were mapped to the same qubit. In our experiments, this did not cumulate too much noise. However, in longer circuits, the cumulative noise would be an issue. For future research, methods from \cite{Meijer_van_de_Griend_2023-1, Meijer_van_de_Griend_2023-2} could be used for the qubit routing to make this method architecture aware, possibly reduce the amount of extra gates inserted and make them more evenly distributed.

Although these optimizations are specific to this algorithm, we feel that other quantum computing researchers could also benefit from the techniques presented in this article. Especially phase folding is a powerful optimization tool whenever there are recurring elements, for example multi-controlled rotation gates, present in the algorithm. The phase folding technique is not implemented in the Qiskit compiler, for example. The technique could be implemented as an optimization routine using the sum-over-paths method presented in \cite{amy2024linearnonlinearrelationalanalyses}, instead of having to do these kinds of optimizations manually.

\section{Conclusions}
In this article, we implemented, optimized, and executed the variational quantum regression algorithm from \cite{wang2024variationalquantumregressionalgorithm} on real quantum hardware. We optimized the regression algorithm using techniques from ZX calculus so that it could run on modern quantum hardware and produce usable regression weights with a decent R2 score on a real-world data set. In the process, we discovered a state preparation method, which could rival the state-of-the-art state preparation methods. Furthermore, the same optimization tricks used in this work could be applied to other algorithms with similar structures.

When executing hybrid algorithms on real hardware, it is important to pay attention to the total gate count, gate and measurement errors, and the choice of optimizer. As we saw with Nelder-Mead and ADAM, some optimizers are more resilient to noise than others. Therefore, it is important to use classical post-processing techniques to help combat the effects of the noise if needed.

For future work, the state preparation method is the most promising for further development. Although we demonstrated that the regression algorithm could be executed on real quantum hardware, classical regression will likely remain dominant for the foreseeable future. However, quantum state preparation is used in many algorithms as a subroutine. Therefore, it would be interesting to fix the problem with the low success probability of the projective measurement. Additionally, the state preparation could be made architecture aware. In that case, the state preparation method would be the new state-of-the-art in terms of overall gate count.

In conclusion, this work highlights the power of hand-crafted optimizations in making quantum algorithms more practical on real hardware. By carefully designing circuits with hardware constraints in mind, we can push the boundaries of what is currently feasible, enabling more complex computations than initially expected. As quantum hardware continues to evolve, such tailored optimizations will play a crucial role in bridging the gap between theoretical algorithms and real-world implementation, bringing us closer to unlocking the full potential of quantum computing.

\appendices
\section{Gate pushing}
\label{ref:app_a}
The gate pushing techniques used in this article are well-known methods from ZX-calculus. In this section, we will showcase illustrative examples of Pauli pushing and Hadamard pushing in terms of quantum circuits and ZX-diagrams. For those more interested in the ZX-calculus rules used, we encourage to refer to \cite{vandewetering2020zxcalculusworkingquantumcomputer} and \cite{Cowtan_2020}.

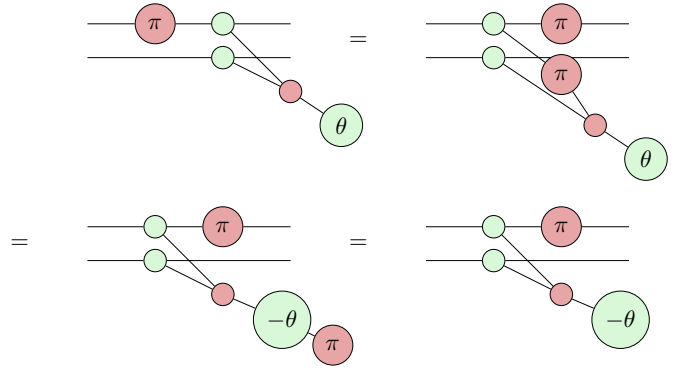
\begin{figure}[!b]
\scalebox{0.9}{
\begin{tikzpicture}
	\begin{pgfonlayer}{nodelayer}
		\node [style=none] (0) at (-1, 1) {};
		\node [style=X Spider] (1) at (1, 1) {$\pi$};
		\node [style=Z Spider] (2) at (3, 1) {};
		\node [style=Z Spider] (3) at (3, 0) {};
		\node [style=none] (4) at (-1, 0) {};
		\node [style=X Spider] (5) at (5, -1) {};
		\node [style=Z Spider] (6) at (6.5, -2) {$\theta$};
		\node [style=none] (7) at (5, 1) {};
		\node [style=none] (8) at (5, 0) {};
		\node [style=none] (9) at (7, 0.5) {$=$};
		\node [style=none] (10) at (9, 1) {};
		\node [style=Z Spider] (12) at (11, 1) {};
		\node [style=Z Spider] (13) at (11, 0) {};
		\node [style=none] (14) at (9, 0) {};
		\node [style=X Spider] (15) at (14, -2) {};
		\node [style=Z Spider] (16) at (15.5, -3) {$\theta$};
		\node [style=none] (17) at (15, 1) {};
		\node [style=none] (18) at (15, 0) {};
		\node [style=X Spider] (19) at (13, 1) {$\pi$};
		\node [style=X Spider] (20) at (13, -0.5) {$\pi$};
		\node [style=none] (21) at (-3, -5.5) {$=$};
		\node [style=none] (22) at (-1, -5) {};
		\node [style=Z Spider] (23) at (1, -5) {};
		\node [style=Z Spider] (24) at (1, -6) {};
		\node [style=none] (25) at (-1, -6) {};
		\node [style=X Spider] (26) at (3, -7) {};
		\node [style=Z Spider] (27) at (4.75, -7.75) {$-\theta$};
		\node [style=none] (28) at (5, -5) {};
		\node [style=none] (29) at (5, -6) {};
		\node [style=X Spider] (30) at (3, -5) {$\pi$};
		\node [style=X Spider] (31) at (6.25, -8.5) {$\pi$};
		\node [style=none] (32) at (7, -5.5) {$=$};
		\node [style=none] (33) at (9, -5) {};
		\node [style=Z Spider] (34) at (11, -5) {};
		\node [style=Z Spider] (35) at (11, -6) {};
		\node [style=none] (36) at (9, -6) {};
		\node [style=X Spider] (37) at (13, -7) {};
		\node [style=Z Spider] (38) at (14.75, -7.75) {$-\theta$};
		\node [style=none] (39) at (15, -5) {};
		\node [style=none] (40) at (15, -6) {};
		\node [style=X Spider] (41) at (13, -5) {$\pi$};
	\end{pgfonlayer}
	\begin{pgfonlayer}{edgelayer}
		\draw (0.center) to (1);
		\draw (1) to (2);
		\draw (2) to (7.center);
		\draw (4.center) to (3);
		\draw (3) to (8.center);
		\draw (3) to (5);
		\draw (2) to (5);
		\draw (5) to (6);
		\draw (13) to (18.center);
		\draw (13) to (15);
		\draw (15) to (16);
		\draw (10.center) to (12);
		\draw (12) to (19);
		\draw (19) to (17.center);
		\draw (14.center) to (13);
		\draw (12) to (20);
		\draw (20) to (15);
		\draw (24) to (29.center);
		\draw (24) to (26);
		\draw (22.center) to (23);
		\draw (23) to (30);
		\draw (30) to (28.center);
		\draw (25.center) to (24);
		\draw (23) to (26);
		\draw (26) to (27);
		\draw (27) to (31);
		\draw (35) to (40.center);
		\draw (35) to (37);
		\draw (33.center) to (34);
		\draw (34) to (41);
		\draw (41) to (39.center);
		\draw (36.center) to (35);
		\draw (34) to (37);
		\draw (37) to (38);
	\end{pgfonlayer}
\end{tikzpicture}
}
\caption{Pushing an X spider through a Z-phase gadget \cite{Cowtan_2020}. This procedure is equivalent to the one in Fig. \ref{fig:zx-circ-example}.}
\label{fig:spiders}
\end{figure}

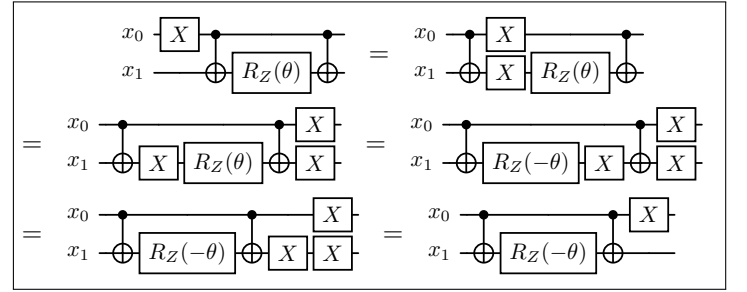
\begin{figure}[!b]
    \centering
        
    \begin{pcvstack}[center, boxed]
    \begin{pchstack}
    \hspace{1cm}
    \scalebox{0.85}{
    \begin{quantikz}[row sep={0.6cm,between origins}, column sep = 0.1cm]
        \lstick{{$x_0$}} & \gate{X} &  \ctrl{1} & & \ctrl{1} & \\
     \lstick{{$x_1$}} &  &\targ{} & \gate{R_Z(\theta)} & \targ{} &
    \end{quantikz}
    }
    $=$
    \scalebox{0.85}{
    
    \begin{quantikz}[row sep={0.6cm,between origins}, column sep = 0.1cm]
        \lstick{{$x_0$}} &  &  \ctrl{1} & \gate{X} & & \ctrl{1} & \\
     \lstick{{$x_1$}} &  &\targ{} & \gate{X} & \gate{R_Z(\theta)} & \targ{} &
    \end{quantikz}
    }
    \end{pchstack}
    
    \begin{pchstack}
    $=$
    \scalebox{0.85}{
    \begin{quantikz}[row sep={0.6cm,between origins}, column sep = 0.1cm]
        \lstick{{$x_0$}} &  &  \ctrl{1} & & & \ctrl{1} & \gate{X} & \\
     \lstick{{$x_1$}} &  &\targ{} & \gate{X} & \gate{R_Z(\theta)}   & \targ{} & \gate{X} &
    \end{quantikz}
    }
    $=$
    \scalebox{0.85}{
    
     \begin{quantikz}[row sep={0.6cm,between origins}, column sep = 0.1cm]
        \lstick{{$x_0$}} &  &  \ctrl{1} & & & \ctrl{1} & \gate{X} & \\
     \lstick{{$x_1$}} &  &\targ{} &  \gate{R_Z(-\theta)} & \gate{X}    & \targ{} & \gate{X} &
    \end{quantikz}
    }
    \end{pchstack}
    \begin{pchstack}
    $=$
    \scalebox{0.85}{
    
     \begin{quantikz}[row sep={0.6cm,between origins}, column sep = 0.1cm]
        \lstick{{$x_0$}} &  &  \ctrl{1} & & \ctrl{1} & &\gate{X} & \\
     \lstick{{$x_1$}} &  &\targ{} &  \gate{R_Z(-\theta)} & \targ{} &\gate{X}& \gate{X} &
    \end{quantikz} 
    }
    $=$
    \scalebox{0.85}{
    
    \begin{quantikz}[row sep={0.6cm,between origins}, column sep = 0.1cm]
        \lstick{{$x_0$}} &  &  \ctrl{1} & & \ctrl{1} &\gate{X} & \\
     \lstick{{$x_1$}} &  &\targ{} &  \gate{R_Z(-\theta)} & \targ{} & &
    \end{quantikz}
    }
    \end{pchstack}
    \end{pcvstack}
    
    \caption{Pushing an X gate through a quantum circuit.}
    \label{fig:zx-circ-example}
\end{figure}

\subsection{Pauli pushing}
\label{ref:app_a-a}
Using the spider fusion rule, the $\pi$-copy rule, and the $\pi$-commute rule from ZX-calculus \cite{vandewetering2020zxcalculusworkingquantumcomputer}, we can push Pauli gates through quantum circuits. In simple terms, pushing an X gate through the control of a CNOT gate will result in X gates on both the control and the target on the other side of the CNOT gate. Pushing an X gate through the target of a CNOT gate is just a matter of swapping the order of the CNOT gate and X gate. Pushing an X gate through an $R_Z$ gate simply changes the sign of the angle. Fig. \ref{fig:zx-circ-example} shows a step-by-step procedure of pushing an X gate through a quantum circuit with $R_Z$ and CNOT gates. For those more familiar with ZX-calculus, Fig. \ref{fig:spiders}  represents the same procedure, illustrated by an X spider and a simple Z-phase gadget.

\begin{figure*}[!h]
    \centering
    \scalebox{1}{
    \begin{quantikz}
         \lstick{$x_0$} & &  \ctrl{1}& & & \ctrl{1} & & & \ctrl{1} & & & \ctrl{1} & & \\
       \lstick{{$x_1$}} & \gate{R_Z(\theta_0 )} &\targ{}  & \wire[l][1]["x_0\oplus x_1"{below,pos=0.1}]{a}&\gate{R_Z(\theta_0 )} & \targ{} & \wire[l][1]["x_1"{below,pos=0.1}]{a}&\gate{R_Z(\theta_1 )} & \targ{} & \wire[l][1]["x_0\oplus x_1"{below,pos=0.1}]{a} &\gate{R_Z(-\theta_1 )} & \targ{} & \wire[l][1]["x_1"{below,pos=0.1}]{a} &
    \end{quantikz}
    }
    \caption{An example circuit.}
    \label{fig:full}
\end{figure*}
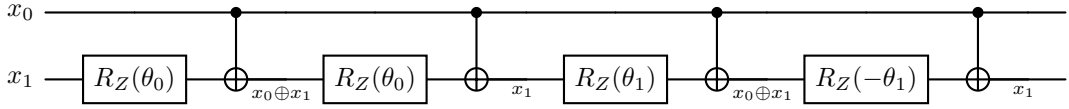

\subsection{Hadamard pushing}
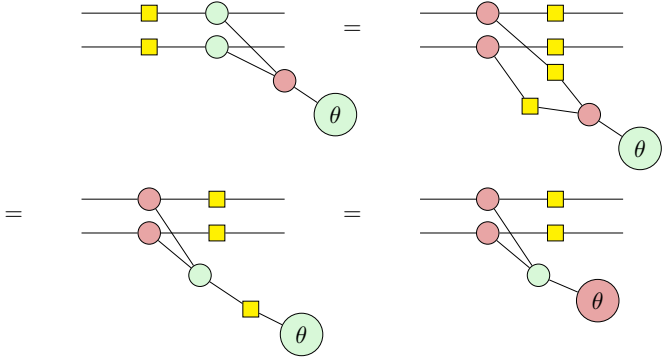
\begin{figure}
\scalebox{0.9}{
\begin{tikzpicture}
	\begin{pgfonlayer}{nodelayer}
		\node [style=none] (0) at (-1, 1) {};
		\node [style=Z Spider] (2) at (3, 1) {};
		\node [style=Z Spider] (3) at (3, 0) {};
		\node [style=none] (4) at (-1, 0) {};
		\node [style=X Spider] (5) at (5, -1) {};
		\node [style=Z Spider] (6) at (6.5, -2) {$\theta$};
		\node [style=none] (7) at (5, 1) {};
		\node [style=none] (8) at (5, 0) {};
		\node [style=none] (9) at (7, 0.5) {$=$};
		\node [style=none] (10) at (9, 1) {};
		\node [style=none] (14) at (9, 0) {};
		\node [style=X Spider] (15) at (14, -2) {};
		\node [style=Z Spider] (16) at (15.5, -3) {$\theta$};
		\node [style=none] (17) at (15, 1) {};
		\node [style=none] (18) at (15, 0) {};
		\node [style=Hadamard] (42) at (1, 1) {};
		\node [style=Hadamard] (43) at (1, 0) {};
		\node [style=Hadamard] (44) at (13, 1) {};
		\node [style=Hadamard] (45) at (13, 0) {};
		\node [style=Hadamard] (46) at (12.25, -1.75) {};
		\node [style=Hadamard] (47) at (13, -0.75) {};
		\node [style=none] (48) at (-1, -4.5) {};
		\node [style=none] (51) at (-1, -5.5) {};
		\node [style=Z Spider] (53) at (5.5, -8.5) {$\theta$};
		\node [style=none] (54) at (5, -4.5) {};
		\node [style=none] (55) at (5, -5.5) {};
		\node [style=Hadamard] (56) at (3, -4.5) {};
		\node [style=Hadamard] (57) at (3, -5.5) {};
		\node [style=none] (60) at (-3, -5) {$=$};
		\node [style=X Spider] (61) at (11, 1) {};
		\node [style=X Spider] (62) at (11, 0) {};
		\node [style=X Spider] (63) at (1, -4.5) {};
		\node [style=X Spider] (64) at (1, -5.5) {};
		\node [style=Z Spider] (65) at (2.5, -6.75) {};
		\node [style=Hadamard] (66) at (4, -7.75) {};
		\node [style=none] (67) at (9, -4.5) {};
		\node [style=none] (68) at (9, -5.5) {};
		\node [style=none] (70) at (15, -4.5) {};
		\node [style=none] (71) at (15, -5.5) {};
		\node [style=Hadamard] (72) at (13, -4.5) {};
		\node [style=Hadamard] (73) at (13, -5.5) {};
		\node [style=none] (74) at (7, -5) {$=$};
		\node [style=X Spider] (75) at (11, -4.5) {};
		\node [style=X Spider] (76) at (11, -5.5) {};
		\node [style=Z Spider] (77) at (12.5, -6.75) {};
		\node [style=X Spider] (78) at (14.25, -7.5) {$\theta$};
	\end{pgfonlayer}
	\begin{pgfonlayer}{edgelayer}
		\draw (2) to (7.center);
		\draw (3) to (8.center);
		\draw (3) to (5);
		\draw (2) to (5);
		\draw (5) to (6);
		\draw (15) to (16);
		\draw (0.center) to (42);
		\draw (42) to (2);
		\draw (4.center) to (43);
		\draw (43) to (3);
		\draw (44) to (17.center);
		\draw (45) to (18.center);
		\draw (46) to (15);
		\draw (47) to (15);
		\draw (56) to (54.center);
		\draw (57) to (55.center);
		\draw (10.center) to (61);
		\draw (61) to (44);
		\draw (14.center) to (62);
		\draw (62) to (45);
		\draw (62) to (46);
		\draw (61) to (47);
		\draw (48.center) to (63);
		\draw (63) to (56);
		\draw (51.center) to (64);
		\draw (64) to (57);
		\draw (63) to (65);
		\draw (64) to (65);
		\draw (65) to (66);
		\draw (66) to (53);
		\draw (72) to (70.center);
		\draw (73) to (71.center);
		\draw (67.center) to (75);
		\draw (75) to (72);
		\draw (68.center) to (76);
		\draw (76) to (73);
		\draw (75) to (77);
		\draw (76) to (77);
		\draw (77) to (78);
	\end{pgfonlayer}
\end{tikzpicture}
}
\caption{Pushing Hadamard gates through a Z-phase gadget \cite{Cowtan_2020}. This procedure is equivalent to the one in Fig. \ref{fig:h-circ-example}.}
\label{fig:hspiders}
\end{figure}

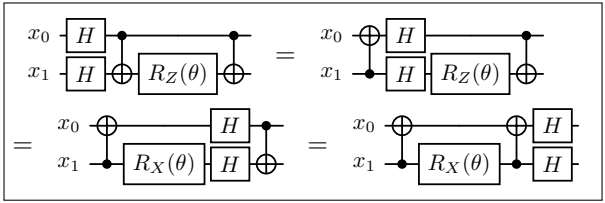
\begin{figure}
    \centering
    \begin{pcvstack}[center, boxed]
    \begin{pchstack}
    \scalebox{0.85}{
    
    \begin{quantikz}[row sep={0.6cm,between origins}, column sep = 0.1cm]
    \lstick{{$x_0$}} & \gate{H} &  \ctrl{1} & & \ctrl{1} & \\
     \lstick{{$x_1$}} & \gate{H} &\targ{} & \gate{R_Z(\theta)} & \targ{} &
    \end{quantikz}
    }
    $=$
    \scalebox{0.85}{
    
    \begin{quantikz}[row sep={0.6cm,between origins}, column sep = 0.1cm]
        \lstick{{$x_0$}}   & \targ{}& \gate{H} & & \ctrl{1} & \\
         \lstick{{$x_1$}}   &\ctrl{-1} & \gate{H} & \gate{R_Z(\theta)} & \targ{} &
    \end{quantikz}
    }
    \end{pchstack}
    \begin{pchstack}
    $=$
    \scalebox{0.85}{
    
\begin{quantikz}[row sep={0.6cm,between origins}, column sep = 0.1cm]
    \lstick{{$x_0$}}   & \targ{}&  &\gate{H} & \ctrl{1} & \\
     \lstick{{$x_1$}}   &\ctrl{-1}  & \gate{R_X(\theta)} & \gate{H} & \targ{} &
    \end{quantikz}
    }
    $=$
    \scalebox{0.85}{
    
    \begin{quantikz}[row sep={0.6cm,between origins}, column sep = 0.1cm]
    \lstick{{$x_0$}}   & \targ{}& & \targ{} & \gate{H} & \\
     \lstick{{$x_1$}}   &\ctrl{-1}  & \gate{R_X(\theta)}  & \ctrl{-1}&\gate{H} &
    \end{quantikz}
    }
    \end{pchstack}
    \end{pcvstack}
    \caption{Pushing Hadamard gates through a quantum circuit.}
    \label{fig:h-circ-example}
\end{figure}

Pushing Hadamard gates is very similar to Pauli pushing. When pushing Hadamard gates, we use the Hadamard identity rule and the Hadamard conjugation rule \cite{vandewetering2020zxcalculusworkingquantumcomputer}. These rules correspond to the well-known circuit identities
\begin{equation}
    HH=I
\end{equation}
and
\begin{equation}
    HZH=X.
\end{equation}
Simply put, when pushing Hadamard gates through a circuit, Z gates are turned into X gates and vice versa. Fig. \ref{fig:h-circ-example} shows a step-by-step procedure
of pushing Hadamard gates through a quantum circuit with $R_Z$ and
CNOT gates. Fig. \ref{fig:hspiders} shows the same procedure with ZX-diagrams.

\section{Phase Folding}
\label{ref:app_b}

Let $C$ be a circuit over $\{X, CNOT, R_Z\}$. Then the circuit can be expressed as
\begin{align}
    [|C|]:|\Vec{x} \rangle \rightarrow e^{2i\pi P(\Vec{x})} |A\Vec{x}+\Vec{b}\rangle,
    \label{eq_2}
\end{align}

where $A\in GL(n, \mathbb{Z}_2)$. In (\ref{eq_2}), $P:\mathbb{Z}_2^n \rightarrow \mathbb{R}$ can be written as

\begin{align}
    P(\Vec{x})= \sum_{\Vec{y}} a_{\Vec{y}}\chi_{\Vec{y}}(\Vec{x}), \hspace{5mm} a_{\Vec{y}} \in \mathbb{R},
\end{align}
where
\begin{align}
    \chi_{\Vec{y}}(\Vec{x})= \Vec{x} \cdot \Vec{y}= x_1y_1 \oplus \dots \oplus x_ny_n, \hspace{5mm} \Vec{x}, \Vec{y} \in \mathbb{Z}_2^n.
\end{align}

$P$ is called the phase polynomial and $(P, A, \Vec{b})$ is called a phase polynomial representation of $C$. The phase polynomial of this kind of circuit can be determined by \textit{annotating} the circuit using the fact that

\begin{align}
    CNOT|x_0, x_1 \rangle = |x_0 \oplus x_1, x_1 \rangle.
\end{align}

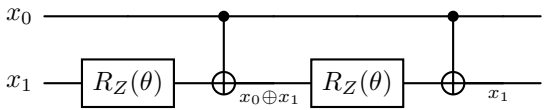
\begin{figure}[!b]
    \centering
    \begin{quantikz}
        \lstick{{$x_0$}} &  &  \ctrl{1}& & & \ctrl{1} & &\\
     \lstick{{$x_1$}} & \gate{R_Z(\theta)} &\targ{} &\wire[l][1]["x_0\oplus x_1"{below,pos=0.1}]{a} &\gate{R_Z(\theta )} & \targ{} & \wire[l][1]["x_1"{below,pos=0.1}]{a} &
    \end{quantikz}
    \caption{An annotated $CR_Z$ circuit.}
    \label{fig:crz}
\end{figure}

An example of an annotated circuit can be seen in Fig. \ref{fig:crz}. The phase polynomial for the circuit in Fig. \ref{fig:crz} would therefore be

\begin{align}
    P(x_0, x_1) = \theta x_1+\theta x_0\oplus x_1.
\end{align}

The $R_Z$ gates which add to the same term in $P$ can be replaced by a single gate. This is called phase folding. As an example, consider the circuit in Fig. \ref{fig:full}. The phase polynomial for that circuit would be
\begin{align}
    P(x_0, x_1) &= \theta_0x_1+\theta_0x_0\oplus x_1+\theta_1x_1 - \theta_1x_0 \oplus x_1 \\
    &= (\theta_0 + \theta_1)x_1 + (\theta_0 - \theta_1)x_0 \oplus x_1.
\end{align}

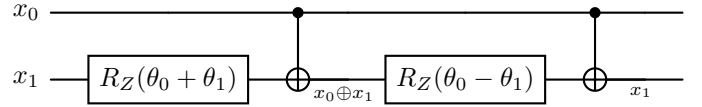
\begin{figure}[!b]
    \centering
    \begin{quantikz}
         \lstick{{$x_0$}} & &  \ctrl{1}& & & \ctrl{1} & &  \\
       \lstick{{$x_1$}} & \gate{R_Z(\theta_0+\theta_1)} &\targ{}  & \wire[l][1]["x_0\oplus x_1"{below,pos=0.1}]{a}&\gate{R_Z(\theta_0-\theta_1)} & \targ{} & \wire[l][1]["x_1"{below,pos=0.1}]{a} & 
    \end{quantikz}
    \caption{An optimized version of the circuit in Fig. \ref{fig:full}.}
    \label{fig:full-optimized}
\end{figure}

We can then replace all the $R_Z$ gates corresponding to the same term in $P$ with a single $R_Z$ gate, yielding the optimized circuit presented in Fig. \ref{fig:full-optimized}. 

\begin{figure}[!b]
\scalebox{0.9}{
\hspace{1cm}
\begin{tikzpicture}
	\begin{pgfonlayer}{nodelayer}
		\node [style=none] (33) at (-2, 1) {};
		\node [style=Z Spider] (34) at (2, 1) {};
		\node [style=Z Spider] (35) at (2, -1) {};
		\node [style=none] (36) at (-2, -1) {};
		\node [style=X Spider] (37) at (4, -3) {};
		\node [style=Z Spider] (38) at (5.75, -3.75) {$\theta_0$};
		\node [style=Z Spider] (39) at (6, 1) {};
		\node [style=Z Spider] (40) at (6, -1) {};
		\node [style=X Spider] (41) at (8, -3) {};
		\node [style=Z Spider] (42) at (9.75, -3.75) {$-\theta_1$};
		\node [style=Z Spider] (43) at (0, -1) {$\theta_0$};
		\node [style=Z Spider] (44) at (4, -1) {$\theta_1$};
		\node [style=none] (45) at (8, -1) {};
		\node [style=none] (46) at (8, 1) {};
		\node [style=none] (47) at (0.75, -6) {};
		\node [style=Z Spider] (48) at (4.75, -6) {};
		\node [style=Z Spider] (49) at (4.75, -8) {};
		\node [style=none] (50) at (0.75, -8) {};
		\node [style=X Spider] (51) at (6.75, -10) {};
		\node [style=Z Spider] (52) at (8.5, -10.75) {\scalebox{0.7}{$\theta_0-\theta_1$}};
		\node [style=Z Spider] (57) at (2.75, -8) {\scalebox{0.7}{$\theta_0+\theta_1$}};
		\node [style=none] (59) at (6.75, -8) {};
		\node [style=none] (60) at (6.75, -6) {};
		\node [style=none] (61) at (-1.25, -7) {$=$};
	\end{pgfonlayer}
	\begin{pgfonlayer}{edgelayer}
		\draw (35) to (37);
		\draw (34) to (37);
		\draw (37) to (38);
		\draw (40) to (41);
		\draw (39) to (41);
		\draw (41) to (42);
		\draw (33.center) to (34);
		\draw [in=180, out=0] (36.center) to (43);
		\draw (43) to (35);
		\draw (34) to (39);
		\draw (35) to (44);
		\draw [in=180, out=0] (44) to (40);
		\draw (40) to (45.center);
		\draw (39) to (46.center);
		\draw (49) to (51);
		\draw (48) to (51);
		\draw (51) to (52);
		\draw (47.center) to (48);
		\draw [in=180, out=0] (50.center) to (57);
		\draw (57) to (49);
		\draw (48) to (60.center);
		\draw (49) to (59.center);
	\end{pgfonlayer}
\end{tikzpicture}
}
\caption{Phase folding using Z-phase gadgets \cite{Cowtan_2020}. }
\label{fig:phase-fold}
\end{figure}
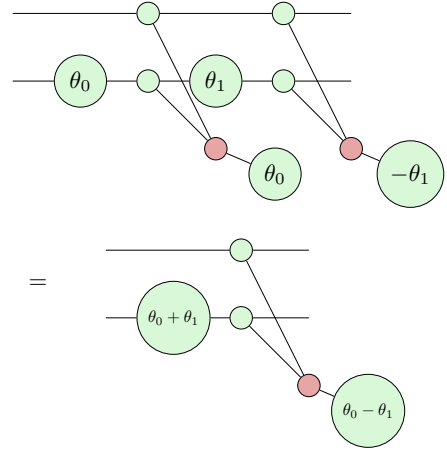

Phase folding becomes very simple when using ZX-calculus. Fig. \ref{fig:phase-fold} shows the phase folding procedure of the circuit in Fig. \ref{fig:full} with ZX-diagrams. In Fig. \ref{fig:phase-fold}, the bi-algebra rule and the spider fusion rule are used \cite{vandewetering2020zxcalculusworkingquantumcomputer}. More rigorous examples of phase folding can be found in \cite{amy2024linearnonlinearrelationalanalyses}.

\newpage
\bibliography{ref}

\begin{thebibliography}{10}

\bibitem{aineisto}
Mohan~S Acharya, Asfia Armaan, and Aneeta~S Antony.
\newblock A comparison of regression models for prediction of graduate admissions.
\newblock In {\em 2019 International Conference on Computational Intelligence in Data Science (ICCIDS)}, pages 1--5, 2019.
\newblock \href {https://doi.org/10.1109/ICCIDS.2019.8862140} {\path{doi:10.1109/ICCIDS.2019.8862140}}.

\bibitem{amy2024linearnonlinearrelationalanalyses}
Matthew Amy and Joseph Lunderville.
\newblock Linear and non-linear relational analyses for quantum program optimization, 2024.
\newblock URL: \url{https://arxiv.org/abs/2410.23493}, \href {https://arxiv.org/abs/2410.23493} {\path{arXiv:2410.23493}}.

\bibitem{Cowtan_2020}
Alexander Cowtan, Silas Dilkes, Ross Duncan, Will Simmons, and Seyon Sivarajah.
\newblock Phase gadget synthesis for shallow circuits.
\newblock {\em Electronic Proceedings in Theoretical Computer Science}, 318:213–228, May 2020.
\newblock URL: \url{http://dx.doi.org/10.4204/EPTCS.318.13}, \href {https://doi.org/10.4204/eptcs.318.13} {\path{doi:10.4204/eptcs.318.13}}.

\bibitem{garnet}
Leonid~Abdurakhimov et.al.
\newblock Technology and performance benchmarks of iqm's 20-qubit quantum computer, 2024.
\newblock URL: \url{https://arxiv.org/abs/2408.12433}, \href {https://arxiv.org/abs/2408.12433} {\path{arXiv:2408.12433}}.

\bibitem{Huang_2020}
Hsin-Yuan Huang, Richard Kueng, and John Preskill.
\newblock Predicting many properties of a quantum system from very few measurements.
\newblock {\em Nature Physics}, 16(10):1050–1057, June 2020.
\newblock URL: \url{http://dx.doi.org/10.1038/s41567-020-0932-7}, \href {https://doi.org/10.1038/s41567-020-0932-7} {\path{doi:10.1038/s41567-020-0932-7}}.

\bibitem{PhysRevA.93.032318}
Raban Iten, Roger Colbeck, Ivan Kukuljan, Jonathan Home, and Matthias Christandl.
\newblock Quantum circuits for isometries.
\newblock {\em Phys. Rev. A}, 93:032318, Mar 2016.
\newblock URL: \url{https://link.aps.org/doi/10.1103/PhysRevA.93.032318}, \href {https://doi.org/10.1103/PhysRevA.93.032318} {\path{doi:10.1103/PhysRevA.93.032318}}.

\bibitem{kingma2017adammethodstochasticoptimization}
Diederik~P. Kingma and Jimmy Ba.
\newblock Adam: A method for stochastic optimization, 2017.
\newblock \href {https://arxiv.org/abs/1412.6980} {\path{arXiv:1412.6980}}.

\bibitem{li2019tacklingqubitmappingproblem}
Gushu Li, Yufei Ding, and Yuan Xie.
\newblock Tackling the qubit mapping problem for nisq-era quantum devices, 2019.
\newblock \href {https://arxiv.org/abs/1809.02573} {\path{arXiv:1809.02573}}.

\bibitem{Meijer_van_de_Griend_2023-2}
Arianne Meijer-van~de Griend and Ross Duncan.
\newblock Architecture-aware synthesis of phase polynomials for nisq devices.
\newblock {\em Electronic Proceedings in Theoretical Computer Science}, 394:116–140, November 2023.
\newblock URL: \url{http://dx.doi.org/10.4204/EPTCS.394.8}, \href {https://doi.org/10.4204/eptcs.394.8} {\path{doi:10.4204/eptcs.394.8}}.

\bibitem{Meijer_van_de_Griend_2023-1}
Arianne Meijer-van~de Griend and Sarah~Meng Li.
\newblock Dynamic qubit routing with cnot circuit synthesis for quantum compilation.
\newblock {\em Electronic Proceedings in Theoretical Computer Science}, 394:363–399, November 2023.
\newblock URL: \url{http://dx.doi.org/10.4204/EPTCS.394.18}, \href {https://doi.org/10.4204/eptcs.394.18} {\path{doi:10.4204/eptcs.394.18}}.

\bibitem{mottonen2004transformationquantumstatesusing}
Mikko Mottonen, Juha~J. Vartiainen, Ville Bergholm, and Martti~M. Salomaa.
\newblock Transformation of quantum states using uniformly controlled rotations, 2004.
\newblock URL: \url{https://arxiv.org/abs/quant-ph/0407010}, \href {https://arxiv.org/abs/quant-ph/0407010} {\path{arXiv:quant-ph/0407010}}.

\bibitem{PRXQuantum.2.040326}
Paul~D. Nation, Hwajung Kang, Neereja Sundaresan, and Jay~M. Gambetta.
\newblock Scalable mitigation of measurement errors on quantum computers.
\newblock {\em PRX Quantum}, 2:040326, Nov 2021.
\newblock URL: \url{https://link.aps.org/doi/10.1103/PRXQuantum.2.040326}, \href {https://doi.org/10.1103/PRXQuantum.2.040326} {\path{doi:10.1103/PRXQuantum.2.040326}}.

\bibitem{vandewetering2020zxcalculusworkingquantumcomputer}
John van~de Wetering.
\newblock Zx-calculus for the working quantum computer scientist, 2020.
\newblock URL: \url{https://arxiv.org/abs/2012.13966}, \href {https://arxiv.org/abs/2012.13966} {\path{arXiv:2012.13966}}.

\bibitem{Vartiainen_2004}
Juha~J. Vartiainen, Mikko Möttönen, and Martti~M. Salomaa.
\newblock Efficient decomposition of quantum gates.
\newblock {\em Physical Review Letters}, 92(17), April 2004.
\newblock URL: \url{http://dx.doi.org/10.1103/PhysRevLett.92.177902}, \href {https://doi.org/10.1103/physrevlett.92.177902} {\path{doi:10.1103/physrevlett.92.177902}}.

\bibitem{wang2024variationalquantumregressionalgorithm}
C.~C.~Joseph Wang and Ryan~S. Bennink.
\newblock Variational quantum regression algorithm with encoded data structure, 2024.
\newblock URL: \url{https://arxiv.org/abs/2307.03334}, \href {https://arxiv.org/abs/2307.03334} {\path{arXiv:2307.03334}}.

\end{thebibliography}
\label{pages:refs}

\end{document}